\documentstyle{article}

\def\text#1{\mbox{#1}}

\newcommand{\be}{\begin{equation}}\newcommand{\ee}{\end{equation}}
\newcommand{\bea}{\begin{eqnarray}}\newcommand{\eea}{\end{eqnarray}}
\newcommand{\bma}[1]{\left(\begin{array}{#1}}
\newcommand{\ema}{\end{array}\right)}
\newcommand{\nn}{\nonumber}
\newcommand{\split}{\nn\\&&\mbox{}}
\newcommand{\eq}[1]{Eq.~(\ref{#1})}

\begin{document}

\author{J.~W.~Bos$^1$, D.~Chang$^{2,1}$, S.~C.~Lee$^1$, Y.~C.~Lin$^2$,\\
  and H.~H.~Shih$^3$
  \\
  {\small ${}^1$Institute of Physics, Academia Sinica, Taipei, Taiwan
    }
  \\
  {\small ${}^2$Department of Physics, National Tsing Hua University,
    Hsinchu, Taiwan }
  \\
  {\small ${}^3$Department of Physics and Astronomy, National Central
    University, Chungli,}\\{\small Taiwan} }

\title{Heavy-baryon chiral perturbation theory and reparametrization
invariance}

\date{}

\maketitle

\begin{abstract}
  We examine the constraints imposed by reparametrization invariance
  on heavy-baryon chiral perturbation theory (HBChPT). We study the
  case of 3 flavors and consider both the strong and weak ($|\Delta S|
  =1$) interaction sector. Some of the parameters in the HBChPT
  Lagrangian are fixed as a consequence of reparametrization
  invariance.  We discuss the consequences for the calculation of
  hyperon weak radiative decays in HBChPT. (hep-ph/9611260).
\end{abstract}

\section{Introduction}

Chiral perturbation theory (ChPT) is an effective field theory for the
description of processes involving hadrons at energies well below the
chiral symmetry breaking scale, $\Lambda_{\chi}$.  The Lagrangian of
ChPT is the most general Lagrangian, in terms of the relevant hadronic
degrees of freedom at that energy scale, consistent with the symmetry
properties of underlying QCD.  In ChPT, amplitudes are expanded in
external momenta and quark masses.  Every term in the Lagrangian
contains an arbitrary coefficient. However, for a process at low
energies only a limited number of these parameters are relevant.  By
comparing to experimental data one can then try to fit these
parameters and, if possible, try to use these fitted values to predict
other experimental observables.

Chiral perturbation theory with baryons is complicated by the baryon
mass in the chiral limit, which is of the same order of magnitude as
$\Lambda_{\chi}$ \cite{gasser}. By using heavy-baryon chiral
perturbation theory (HBChPT) \cite{hbcpt} this complication can be
dealt with.  As was pointed out by Ecker and Moj\v{z}i\v{s}
\cite{ecker} a general Lagrangian constructed starting directly in the
heavy-baryon formulation is due to the presence of the external
four-velocity $v^\mu$ not necessarily Lorentz invariant, or
equivalently, not necessarily reparametrization invariant
\cite{luke,dugan}.  Reparametrization invariance constrains the number
of independent terms in an effective Lagrangian \cite{luke,chen}.  In
this paper we will study the constraints of reparametrization
invariance on the general HBChPT Lagrangian considered in
Ref.~\cite{OP95}.  Reparametrization invariance of the HBChPT
Lagrangian can be ensured by matching it with the fully relativistic
Lagrangian \cite{bernard}. This approach we will follow in this paper.

This paper is organized as follows. In the next section we will start
with the relativistic ChPT Lagrangian in both the strong and weak
interaction sectors, and consider the case of 3 flavors.  In Sect. 3
we will then use a path integral method to arrive at the corresponding
HBChPT Lagrangian. We will compare the result with the HBChPT
Lagrangian, derived earlier. In Sect.  4 we will discuss the
consequences of the constraints on the analysis of hyperon weak
radiative decay.  Finally, Sect. 5 contains a summary and our
conclusions.

\section{Construction of the Lagrangian}

In this section we will study the fully relativistic Lagrangian of
ChPT with 3 flavors in both the strong and weak interaction sectors.
The strong Lagrangian can be expanded as
\be
\label{rel_lag}
{\cal L}_s={\cal L}^{(1,0)}_s+{\cal L}^{(0,1)}_s+{\cal L}^{(2,0)}_s
+{\cal L}^{(1,1)}_s+{\cal L}^{(0,2)}_s+\ldots 
\ee 
where the superscript $(i,j)$ denotes that the Lagrangian is of order
$q^i$ in the momentum expansion and of order $m_s^j$ in the mass
expansion.  (For simplicity we take $m_u=m_d=0$.) This Lagrangian has
been analyzed before by Krause \cite{krause}.  In his analysis Krause
takes $m_s$ and $q^2$ to be of the same order in the chiral expansion,
as is customary in the meson sector.  However, arguments below will
indicate that in the baryon sector the relative counting between $q$
and $m_s$ might be different.  Since we use a different expansion
systematic we will again go through all the terms in the Lagrangian.
The Lagrangian ${\cal L}_s^{(1,0)}$ in \eq{rel_lag} can be readily
constructed. Using the same notation for the fields as Krause, it is
given by
\be
\label{rel_lag_10}
{\cal L}_s^{(1,0)}=\langle\bar{B}i\gamma^\mu[D_\mu,B]\rangle-
\dot{m}\langle\bar{B}B\rangle+D\langle\bar{B}i\gamma^5
\gamma^\mu\{\Delta_\mu,B\}\rangle+F\langle\bar{B}i\gamma^5
\gamma^\mu[\Delta_\mu,B]\rangle,
\ee
where $\dot{m}$ is the baryon mass in the chiral-SU(3) limit and $\langle
\ldots\rangle$ denotes a trace in flavor space.  The parameters $D$
and $F$ can be fitted to experimental data.  For example, Jenkins and
Manohar \cite{DFfit} found by fitting to baryon semi-leptonic decays
the values $D=0.61$ and $F=0.40$.
 
Baryons described by ${\cal L}_s^{(1,0)}$ satisfy the equations of
motion
\be
\label{EOM}
i\gamma^\mu[D_\mu,B] - \dot{m}B + D i\gamma^5\gamma^\mu\{\Delta_\mu,
B\} + F i\gamma^5\gamma^\mu [\Delta_\mu,B] - \frac{2}{3}Di\gamma^5\gamma^\mu
\langle\Delta_\mu B\rangle = 0,
\ee
which can be rewritten as
\be
\label{10_EOM}
i\partial\!\!\!\slash B(x) = \dot{m} B(x) + \dots,
\ee
where the dots denote terms that are at least of order $O(q)$.  Therefore,
a derivative of the baryon field counts as order
$O(\dot{m}/\Lambda_\chi) = O(1)$. In particular, the covariant
derivative acting on $B$, $[D^\mu,B] = \partial^\mu B+ [\Gamma^\mu,B]$,
is also of order $O(1)$.  Nevertheless, from \eq{10_EOM} the
combination
\be
\langle\bar{B}i\gamma^\mu[D_\mu,B]\rangle-\dot{m}\langle\bar{B}B\rangle
\ee
appearing in ${\cal L}_s^{(1,0)}$, is of order $O(q)$.  The fact that
$[D^\mu,B]$ is of order $O(1)$ raises an apparent complication in the
expansion of the general ChPT Lagrangian.  One would expect that terms
of the form
\be
\label{EXAMPLE1}
\langle\bar{B}\Gamma^{\mu_1\ldots\mu_n}[D_{\mu_1}, \ldots
,[D_{\mu_n},B]\ldots]\rangle, 
\ee 
where $\Gamma^{\mu_1\ldots\mu_n}$ is some arbitrary operator in Dirac
space, are of order $O(1)$.  Therefore, it seems impossible to expand
the general Lagrangian with a finite number of terms in each order.
However, it can be shown \cite{krause} that most terms with covariant
derivatives acting on $B$ can be written as a linear combination of
terms with {\em less} covariant derivatives acting on $B$ and terms
that are proportional to the equation of motion (\ref{EOM}).  Terms in
an effective Lagrangian that are proportional to the classical
equations of motion can be eliminated by a suitable field redefinition
\cite{FT} (See also Ref.~\cite{stefan} for a discussion on the use of
field redefinitions in mesonic ChPT).  The Lagrangians before and
after such a field redefinition will give the same predictions for any
physical amplitude.  In the weak interaction sector we will consider
an example of how a field transformation can eliminate a term from the
Lagrangian.

By using such field redefinitions, it is then possible to cast the
general Lagrangian in a form in which each order in its expansion
contains a finite number of terms.  Of course, for a practical
calculation of an amplitude in ChPT it is convenient to write the
Lagrangian in a form with the least number of independent terms in
each order. In the following we will attempt to give such a minimal
set of terms. To do so we will also make use of the Cayley identities,
which can relate terms in the Lagrangian with a double trace in flavor
space to terms with a single trace in flavor space.

In this way, we have found that the Lagrangians ${\cal L}_s^{(0,1)}$,
${\cal L}_s^{(2,0)}$, ${\cal L}_s^{(1,1)}$, and ${\cal L}_s^{(0,2)}$
consist of the terms
\bea
\label{rel_lag_01}
\langle\bar{B}(\sigma,B)\rangle;\;
\langle\bar{B}B\rangle\langle\sigma\rangle,
\eea
\bea
\label{rel_lag_20}
&&\langle\bar{B}i\sigma^{\mu\nu}([D_\mu,D_\nu],B)\rangle;
\;
\langle\bar{B}i\sigma^{\mu\nu}([\Delta_\mu,\Delta_\nu],B)\rangle;
\;
\langle\bar{B}(\Delta^\mu,(\Delta_\mu,B))\rangle;
\;
\langle\bar{B}i\gamma^\mu(\Delta_\mu,(\Delta^\nu,[D_\nu,B]))\rangle;
\split
\langle\bar{B}\Delta^\mu\rangle i\sigma_{\mu\nu}\langle\Delta^\nu B\rangle;
\;
\langle\bar{B}\Delta^\mu\rangle\langle\Delta_\mu B\rangle;
\;
\langle\bar{B}\Delta^\mu\rangle i\gamma_\mu \langle\Delta^\nu
[D_\nu,B]\rangle,
\eea
\bea
\label{rel_lag_11}
&&\langle \bar{B}\gamma^5(\rho,B)\rangle;
\;
\langle \bar{B}\gamma^5B\rangle\langle\rho\rangle;
\;
\langle \bar{B}i\gamma^5\gamma^\mu(\Delta_\mu,B)\rangle\langle\sigma
\rangle;
\;
\langle \bar{B}\Delta^\mu\rangle i\gamma^5\gamma_\mu\langle\sigma B\rangle;
\;
\langle \bar{B}\Delta^\mu\rangle\gamma_\mu\langle\rho B\rangle;
\;
\split
\langle\bar{B}i\gamma^5\gamma^\mu(\Delta_\mu,(\sigma,B))\rangle;
\;
\langle\bar{B}i\gamma^\mu([\Delta_\mu,\rho],B)\rangle,
\eea
and
\bea
&&\langle\bar{B}(\sigma,(\sigma,B))\rangle;
\;
\langle\bar{B}(\rho,(\rho,B))\rangle;
\;
\langle\bar{B}B\rangle\times\langle\rho^2\rangle;
\; 
\langle\bar{B}B\rangle\times\langle\sigma^2\rangle;
\;
\langle\bar{B}(\sigma,B)\rangle\times\langle\sigma\rangle;
\;
\langle\bar{B}(\rho,B)\rangle\times\langle\rho\rangle;
\;
\split
\langle\bar{B}\sigma\rangle\times\langle\sigma\,B\rangle;
\; 
\langle\bar{B}\rho\rangle\times\langle\rho B\rangle,
\eea
respectively. In the above compact notation $(A,B)$ can be the
anti-commutator $\{A,B\}$ or the commutator $[A,B]$.  Furthermore,
$(A,(B,C))$ can be any of the combinations $\{A,\{B,C\}\}$,
$\{A,[B,C]\}$, $[A,\{B,C\}]$, or $[A,[B,C]]$. (It is easy to show that
$(A,(A,C))$ represents only three independent combinations.)
Furthermore, a term $X$ in one of the above lists is an abbreviations
for the combination $(X+X^c)/2$, where $X^c$ is the charge conjugated
of $X$.  From the above expressions it an be easily seen that the
Lagrangian ${\cal L}_s^{(0,1)}$ consists of 3 independent terms, the
Lagrangian ${\cal L}_s^{(2,0)}$ consists of 13 independent terms, the
Lagrangian ${\cal L}_s^{(1,1)}$ consists of 13 independent terms, and
the Lagrangian ${\cal L}_s^{(0,2)}$ consists of 14 independent terms.

The Lagrangian ${\cal L}_s^{(0,1)}$ breaks SU(3) symmetry. By
including SU(3) breaking through ${\cal L}_s^{(0,1)}$ the equation of
motion for free octet baryons reads
\be
\label{EOM_SU3}
(i\partial\!\!\!\slash - \dot{m}) B(x) = \Delta m \,B(x),
\ee
where $\Delta m$ is a quantity of order $O(m_s)$, determined by the
parameters in \eq{rel_lag_01}.  As is discussed after \eq{EOM}, the
left-hand side of \eq{EOM_SU3} counts as order $O(q)$ in the chiral
expansion. This indicates that $m_s$ is of order $q$, rather than
$q^2$ as in the meson sector. We again emphasize that in principle in
ChPT $m_s$ and $q$ should be considered as separate expansion
parameters, and only phenomenological analysis will dictate the
relative order.

Necessary ingredients in the general weak ($|\Delta S|=1$) interaction
Lagrangian are the fields $\lambda$ and $\lambda'$ defined e.g.\ in
Ref.~\cite{OP95}.  In the construction of the weak interaction
Lagrangian one encounters the same problem as in the strong
interaction sector, namely that $[D^\mu,B]$ counts as order $O(1)$ in
the chiral expansion.  However, also in the weak interaction sector
this problem can be dealt with by redefining the baryon field.  For
example, the term
\be
\label{EXAMPLE2}
\alpha\Bigl[\langle\bar{B}i\gamma^5\gamma^\mu[\lambda,[D_\mu,B]]\rangle+
\langle\bar{B}i\gamma^5\gamma^\mu[D_\mu,[\lambda,B]]\rangle\Bigr],
\ee
can be effectively
removed from the Lagrangian by the field redefinition
\be
\label{TRANS}
B\rightarrow B + \alpha \gamma^5 [\lambda,B].
\ee
Under the redefinition (\ref{TRANS}) the kinetic part of the strong
Lagrangian (\ref{rel_lag_10} will generate a term that cancels with
\eq{EXAMPLE2}.  The weak interaction Lagrangian can be readily
obtained.  Its expansion reads
\be
{\cal L}_w={\cal L}_w^{(0,0)}+{\cal L}_w^{(1,0)}+{\cal L}_w^{(0,1)}
+\ldots .
\ee
where ${\cal L}_w^{(0,0)}$ is given by
\be
{\cal L}_w^{(0,0)}=h_D\langle\bar{B}\{\lambda,B\}\rangle+
h_F\langle\bar{B}[\lambda,B]\rangle,
\ee
${\cal L}_w^{(1,0)}$ consists of the terms
\bea
\label{WLAG10}
&&\langle\bar{B}i\gamma^5(\lambda',B)\rangle;
\;
\langle\bar{B}i\gamma^\mu(\lambda,(\Delta_\mu,B))\rangle;
\;
\langle\bar{B}i\gamma^5\gamma^\mu(\lambda,(\Delta_\mu,B))\rangle;
\;
\langle\bar{B}i\gamma^\mu([\lambda',\Delta_\mu],B)\rangle;
\;
\langle\bar{B}i\gamma^5\gamma^\mu([\lambda',\Delta_\mu],B)\rangle;
\;
\split
\langle\bar{B}\lambda\rangle i\gamma^\mu\langle\Delta_\mu B\rangle;
\;
\langle\bar{B}\lambda' \rangle i\gamma^\mu\langle\Delta_\mu B\rangle;
\;
\langle\bar{B}\lambda \rangle i\gamma^5\gamma^\mu\langle\Delta_\mu B\rangle;
\;
\langle\bar{B}\lambda'\rangle i\gamma^5\gamma^\mu\langle\Delta_\mu B,\rangle;
\;
\langle\bar{B}\lambda'\rangle\langle\Delta^\mu[D_\mu,B]\rangle,
\eea
and ${\cal L}_w^{(0,1)}$ consists of the terms
\bea
&&\langle\bar{B}(\lambda,(\sigma,B))\rangle;
\;
\langle\bar{B}i(\lambda',(\rho,B))\rangle;
\;
\langle\bar{B}i([\lambda',\sigma],B)\rangle;
\;
\langle\bar{B}([\lambda,\rho],B)\rangle;
\;
\langle\bar{B}(\lambda,B)\rangle\times\langle\sigma\rangle;
\;
\langle\bar{B}i(\lambda',B)\rangle\times\langle\rho\rangle;
\;
\split
\langle\bar{B}\lambda\rangle\times\langle\sigma B\rangle;
\;
\langle\bar{B}\lambda'\rangle\times i\times\langle\rho B\rangle;
\;
\langle\bar{B}\lambda'\rangle\times i\times\langle\sigma B\rangle;
\;
\langle\bar{B}\lambda\rangle\times\langle\rho B\rangle.
\eea
A term $X$ in the above lists for ${\cal L}_w^{(0,1)}$ and ${\cal
L}_w^{(1,0)}$ is an abbreviation for $X+X^{cp}$, where $X^{cp}$ is $X$
after a charge conjugation and a parity operation.  Note that in some
cases the chiral order of a given term is not so obvious.
For example, the list for ${\cal L}_w^{(1,0)}$ does
not contain the term
\be
X = \langle\bar{B}\lambda\rangle\langle\Delta^\mu[D_\mu,B]\rangle,
\ee
since $X+X^{cp}$, given
by
\be
X+X^{cp}=\langle\bar{B}\lambda\rangle\langle\Delta^\mu[D_\mu,B]\rangle+
\langle[\bar{B},D_\mu]\lambda\rangle\langle\Delta^\mu B\rangle,
\ee
can be shown to be of order $O(q^2)$ by using Cayley's identity.  In
conclusion, we have found that the weak Lagrangians ${\cal
  L}_w^{(0,0)}$, ${\cal L}_w^{(1,0)}$, and ${\cal L}_w^{(0,1)}$
consist of 2, 19, and 20 independent terms, respectively.

\section{The non-relativistic limit}

To obtain the non-relativistic limit of the above Lagrangian we use the
path integral approach of Ecker and Moj\v{z}i\v{s} \cite{ecker}, and
Bernard {\em et al.} \cite{bernard}, who studied the case of 2
flavors.  The method is similar to the approach used in heavy-quark
theory by Mannel {\em et al.\/} \cite{mannel}.  The first step consist of
writing the Lagrangian in terms of the velocity dependent fields
\be
H(v,x)={\rm e}^{i\dot{m}v\cdot x}P^+_vB(x),\;\; 
h(v,x)={\rm e}^{i\dot{m}v\cdot x}P^-_vB(x),
\ee
where $v$ is a four-velocity satisfying $v^2 = 1$, and $P_v^+$ and
$P_v^-$ are the velocity dependent projection operators
\be
P^\pm_v=\frac{1\pm v\!\!\!\slash}{2}.
\ee
By expanding the baryon field $B$ as
\be
B=\frac{1}{\sqrt{2}}B^i \lambda^i
\ee
with $\lambda^i$ the Gell-Mann matrices (or any basis will do for that
matter), it is straightforward to extend the formalism of Ecker and
Moj\v{z}i\v{s} and Bernard {\em et al.} to 3 flavors.  (Recently, we
have received a preprint \cite{mueller} in which in a similar way the
case of 2 flavors has been extended to 3 flavors.)  After the above
field redefinitions the Lagrangian (\ref{rel_lag}) is given by
\be
{\cal L}_s = \bar{H}^jM_1^{ji}H^i+\bar{h}^jM_2^{ji}H^i
+\bar{H}^j\gamma_0 (M_2^\dagger)^{ji}\gamma_0h^i
-\bar{h}^jM_3^{ji}h^i,
\ee
where $M_1$, $M_2$ and $M_3$ are $8\times 8$ matrices given
by
\be
2M_1^{ji}=iv^\mu\langle\lambda_j[ D_\mu,\lambda_i]\rangle
-2iDS_v^\mu\langle\lambda_j\{\Delta_\mu,\lambda_i\}\rangle
-2iFS_v^\mu\langle\lambda_j[\Delta_\mu,\lambda_i]\rangle+\ldots,
\ee
\bea
2M_2^{ji}&=&P^-_v\Bigl[
i\gamma^\mu\langle\lambda_j[D_\mu,\lambda_i]\rangle
+iD\gamma^5\langle\lambda_j\{v\cdot\Delta,\lambda_i\}\rangle
+iF\gamma^5\langle\lambda_j[v\cdot\Delta,\lambda_i]\rangle
\split
+c_1\gamma_5 \delta^{ji} \langle\rho\rangle
+c_2\gamma_5 \langle\lambda_j \{\rho,\lambda_i\}\rangle
+c_3\gamma_5 \langle\lambda_j [\rho,\lambda_i]\rangle
+\dots\Bigr]P^+_v,
\eea
and
\bea
M_3^{ji}=2\dot{m}\delta^{ji}+\ldots\;,
\eea
with $S^\mu_v$ is the spin operator defined by
\bea
S_v^\mu\equiv-\frac{1}{2}P^+_v\gamma_5\gamma^\mu P^+_v,
\eea
and $c_1$, $c_2$, and $c_3$, are constants from the first two terms in
Lagrangian (\ref{rel_lag_11}).  The dots in the expressions for $M_2$
and $M_3$ represent terms that are not relevant for the present
analysis.

After the field redefinition
\be
h^i\rightarrow h^i-(M_3^{-1}M_2)^{ij}H^j
\ee
the minus component field $h$ can be easily integrated out from the
Lagrangian, using the path integral formalism. The resulting strong
Lagrangian ${\cal L}_{s,v}$, which should give equivalent physical
prediction as ${\cal L}_{s}$, reads
\be
\label{NRLAG}
{\cal L}_{s,v}=\bar{H}^jM_1^{ji}H^i+\bar{H}^j(\gamma_0M_2^\dagger
\gamma_0M_3^{-1}M_2)^{ji}H^i.
\ee
Using
\be
(M_3^{-1})^{ij} = \frac{1}{2\dot{m}}\delta^{ij} + \ldots
\ee
it is straightforward to expand ${\cal L}_{s,v}$ in powers of $q$ and
$m_s$ (note that each covariant derivatives acting on $H$ counts
as order $q$). The expansion of ${\cal L}_{s,v}$ reads
\be
{\cal L}_{s,v}={\cal L}^{(1,0)}_{s,v}+{\cal L}^{(0,1)}_{s,v}+
{\cal L}^{(2,0)}_{s,v}+{\cal L}^{(1,1)}_{s,v}+
{\cal L}^{(0,2)}_{s,v}+\ldots\;.
\ee
where ${\cal L}^{(1,0)}_{s,v}$ is given by
\bea
{\cal L}^{(1,0)}_{s,v}=\langle\bar{H}i[v\cdot D,H]\rangle
-2iD\langle\bar{H}S_v^\mu\{\Delta_\mu\;,H\}\rangle
-2iF\langle\bar{H}S_v^\mu[\Delta_\mu\;,H]\rangle,
\eea
and
${\cal L}^{(0,1)}_{s,v}$ consists of the terms
\bea
\langle\bar{H}(\sigma,H)\rangle;\,
\langle\bar{H}H\rangle\langle\sigma\rangle.
\eea
The Lagrangian ${\cal L}^{(2,0)}_{s,v}$ is given by
\bea
\label{nrel_lag_20}
{\cal L}^{(2,0)}_{s,v}&=&\frac{1}{2\dot{m}}\langle\bar{H}
[v\cdot D,[v\cdot D,H]]\rangle
-\frac{1}{2\dot{m}}\langle\bar{H}
[D^\mu,[D_\mu,H]]\rangle
\split
-\frac{D}{\dot{m}}\langle\bar{H} S_v^\mu
[D_\mu,\{v\cdot\Delta, H\}]+\bar{H}S_v^\mu
\{v\cdot\Delta,[D_\mu,H]\}\rangle
\split
-\frac{F}{\dot{m}}\langle\bar{H}S_v^\mu
[D_\mu,[v\cdot\Delta,H]]+\bar{H}S_v^\mu
[v\cdot\Delta,[D_\mu,H]]\rangle + \Delta{\cal L}^{(2,0)}_{s,v},
\eea
where $D$ and $F$ are the same constants as in Lagrangian (\ref{rel_lag_10}),
and $\Delta{\cal L}^{(2,0)}_{s,v}$ consists of the terms
\bea
&&\langle\bar{H}[S_v^\mu,S_v^\nu]([D_\mu,D_\nu],H)\rangle;
\;
\langle\bar{H}[S_v^\mu,S_v^\nu]([\Delta_\mu,\Delta_\nu],H)\rangle;
\;
\langle\bar{H}(\Delta^\mu,(\Delta_\mu,H))\rangle;
\;
\langle\bar{H}(v\cdot\Delta,(v\cdot\Delta,H))\rangle;
\;
\split
\langle\bar{H}\Delta^\mu\rangle\times\langle\Delta_\mu H\rangle;
\;
\langle\bar{H}\Delta_\mu\rangle\times[S^\mu_v,S^\nu_v]\times
\langle\Delta_\nu H\rangle;
\;
\langle\bar{H}v\cdot\Delta\rangle\times\langle v\cdot\Delta H\rangle.
\eea
The Lagrangian ${\cal L}^{(1,1)}_{s,v}$ consists of the terms
\bea
&&\langle\bar{H}S_v^\mu(\sigma,(\Delta_\mu,H))\rangle;
\;
\langle\bar{H}([v\cdot\Delta,\rho],H)\rangle;
\;
\langle\bar{H}S_v^\mu(\Delta_\mu,H)\rangle\times\langle\sigma\rangle;
\;
\langle\bar{H}\Delta_\mu\rangle\times S^\mu_v\times\langle\sigma H\rangle;
\;
\split
\langle\bar{H}v\cdot\Delta\rangle\times\langle\rho H\rangle;
\;
\langle\bar{H}S_v^\mu([D_\mu,\rho],H)\rangle;
\;
\langle\bar{B}S_v^\mu B\rangle\langle[D_\mu,\rho]\rangle,
\eea
and the Lagrangian ${\cal L}^{(0,2)}_{s,v}$ consists of the terms 
\be
\langle\bar{H}(\sigma,(\sigma,H))\rangle;
\;
\langle\bar{H}(\rho,(\rho,H))\rangle;
\;
\langle\bar{H}(\sigma,H)\rangle\times\langle\sigma\rangle;
\;
\langle\bar{H}(\rho,H)\rangle\times\langle\rho\rangle;
\;
\langle\bar{H}\sigma\rangle\times\langle\sigma\,H\rangle;
\; 
\langle\bar{H}\rho\rangle\times\langle\rho H\rangle.
\ee
Note that the terms
\be
\langle\bar{H}S_v^\mu([D_\mu,\rho],H)\rangle;
\;
\langle\bar{B}S_v^\mu B\rangle\langle[D_\mu,\rho]\rangle,
\ee
enter ${\cal L}^{(1,1)}_{s,v}$ through the $1/m$ expansion, i.e.,
through the second term in the right-hand side of \eq{NRLAG}.

To obtain the non-relativistic weak Lagrangian the same steps can be
followed. To obtain the $1/m$ part for the weak Lagrangian we need
\bea
M_2^{ji}({\rm weak})&=&\frac{1}{2}P^-_v\Bigl[d_1\gamma^5
\langle\lambda_j\{\lambda',\lambda_i\}\rangle
+d_2\gamma^5
\langle\lambda_j[\lambda',\lambda_i]\rangle+\ldots\Bigr]P^+_v,
\eea
where $d_1$ and $d_2$ are constants corresponding to the first term in
Lagrangian (\ref{WLAG10}), and the dots denote terms at least
containing one of the fields $D^\mu$, $\Delta^\mu$, $\rho$ or
$\sigma$.  The resulting non-relativistic weak Lagrangian, obtained in
the same way as \eq{NRLAG}, then reads
\be
{\cal L}_{w,v}={\cal L}^{(0,0)}_{w,v}+
{\cal L}^{(1,0)}_{w,v}+{\cal L}^{(0,1)}_{w,v}+\ldots,
\ee
where the Lagrangian ${\cal L}_{w,v}^{(0,0)}$ is given by
\be
{\cal L}_{w,v}^{(0,0)} = h_D \langle\bar{H}\{\lambda,H\}\rangle+
h_F\langle\bar{H}[\lambda,H]\rangle,
\ee
the Lagrangian ${\cal L}_{w,v}^{(1,0)}$ consists of the terms
\bea
&&\langle\bar{H}i(\lambda,(v\cdot\Delta,H))\rangle;
\;
\langle\bar{H}iS_v^\mu(\lambda,(\Delta_\mu,H))\rangle;
\;
\langle\bar{H}i([\lambda',v\cdot\Delta],H)\rangle;
\;
\split
\langle\bar{H}iS_v^\mu([\lambda',\Delta_\mu],H)\rangle;
\;
\langle\bar{H}i([v\cdot D,\lambda'],H)\rangle;
\;
\langle\bar{H}iS_v^\mu([D_\mu,\lambda'],H)\rangle;
\;
\langle\bar{H}\lambda\rangle i\langle v\cdot\Delta H\rangle;
\split
\;
\langle\bar{H}\lambda'\rangle i\langle v\cdot\Delta H\rangle;
\;
\langle\bar{H}\lambda\rangle iS_v^\mu\langle\Delta_\mu H\rangle;
\;
\langle\bar{H}\lambda'\rangle iS_v^\mu\langle\Delta_\mu H\rangle,
\eea
and the Lagrangian ${\cal L}_{w,v}^{(0,1)}$ consists of the terms
\bea
&&\langle\bar{H}(\lambda,(\sigma,H))\rangle;
\;
\langle\bar{H}(\lambda',(\rho,H))\rangle;
\;
\langle\bar{H}([\lambda',\sigma],H)\rangle;
\;
\langle\bar{H}([\lambda,\rho],H)\rangle;
\;
\langle\bar{H}(\lambda,H)\rangle\times\langle\sigma\rangle;
\;
\langle\bar{H}(\lambda',H)\rangle\times\langle\rho\rangle;
\split
\;
\langle\bar{H}\lambda\rangle\times\langle\sigma H\rangle;
\;
\langle\bar{H}\lambda'\rangle\times i\times\langle\rho H\rangle;
\;
\langle\bar{H}\lambda'\rangle\times i\times\langle\sigma H\rangle;
\;
\langle\bar{H}\lambda\rangle\times\langle\rho H\rangle.
\eea

The above results can be compared with the results for the general
Lagrangian given in Ref.~\cite{OP95}, which was derived by starting
directly in heavy-baryon formulation of ChPT.  This makes it possible
to derive the constraints imposed by reparametrization invariance on
the HBChPT Lagrangian.  Using \eq{nrel_lag_20} one can easily see
that as an result of reparametrization invariance the coefficients of
the strong interaction terms
\bea
&&\langle\bar{H}[D^\mu,[D_\mu,H]]\rangle;
\;
\split
\langle\bar{H}[v\cdot D,[v\cdot D,H]]\rangle;
\;
\split
\langle\bar{H} S_v^\mu
[D_\mu,\{v\cdot\Delta, H\}]+\bar{H}S_v^\mu
\{v\cdot\Delta,[D_\mu,H]\}\rangle;
\split
\langle\bar{H}S_v^\mu
[D_\mu,[v\cdot\Delta,H]]+\bar{H}S_v^\mu
[v\cdot\Delta,[D_\mu,H]]\rangle,
\eea
in Ref.~\cite{OP95}
are given by $-1/(2m)$,
$1/(2m)$, $-D/m$, $-F/m$, respectively.
In the weak interaction sector, the coefficients of the terms
\bea
\label{aap}
&&\langle\bar{H}S^\mu[D_\mu,\{\lambda,H\}]\rangle+
\langle\bar{H}S^\mu\{\lambda,[D_\mu,H]\}\rangle;\;
\split
\langle\bar{H}S^\mu[\lambda,[D_\mu,H]]\rangle+
\langle\bar{H}S^\mu[D_\mu,[\lambda,H]]\rangle
\eea
in Ref.~\cite{OP95}
are both zero. All other terms in the HBChPT Lagrangian derived in
Ref.~\cite{OP95} are not constrained by reparametrization invariance.

\section{Application to hyperon weak radiative decays}

As discussed in the previous section some of the coefficients in the
general HBChPT Lagrangian are constrained by reparametrization
invariance.  In particular, the coefficients of the weak-interaction
terms in \eq{aap} are constrained to be zero.  In Ref.~\cite{OP96} it
was shown that these constants, denoted there by $a_5$ and $a_6$, are
crucial for the description of the two charged weak radiative decay
channels,
\be
\Xi^- \rightarrow\Sigma^-+\gamma,
\ee
and
\be
\Sigma^+ \rightarrow p+\gamma.
\ee
In leading order, the parity-violating amplitudes ($B$) for the
charged channels were both shown to be proportional to a linear
combination of $a_5$ and $a_6$, while the parity-conserving amplitudes
($A$) for the charged channels were found to be zero.  By fitting the
constants $a_5$ and $a_6$ is was possible to give a satisfactory
description of the decay rates of the charged channels.  However, from
the requirement that $a_5$ and $a_6$ are fixed to be zero it follows
that in leading order not only $A$, but also $B$ for the two charged
channels is vanishing.  This result is in agreement with Jenkins {\em
  et al.} \cite{HRD} who argue that the amplitude $B$ for the charged
channels cannot receive any short distance contributions.  Therefore,
to obtain a non-zero value for both $A$ and $B$ one needs to go at
least to the next-to-leading order.  In such analysis one should also
consider loop-diagrams contributions to the decay as studied by
Refs.~\cite{HRD,neufeld}.

In Ref.~\cite{OP96} it was shown that Hara's theorem, which states
that $B=0$ in the SU(3) symmetric limit, was violated if $a_5$ and
$a_6$ are non-zero.  Since $a_5$ and $a_6$ must be zero by
reparametrization invariance, we conclude that Hara's theorem, is
still satisfied in HBChPT.  In the past, the measured value of
asymmetry parameter for the $\Sigma^+$ decay ($\alpha = -0.75 \pm
0.08$), has been considered as being inconsistent with Hara's theorem.
Indeed, if $B=0$ and $A$ is non-zero, the asymmetry parameter defined
by
\be
\alpha = \frac{{\rm Re} (AB^{*})}{|A|^2 + |B|^2},
\ee
is vanishing. However, in HBChPT both $A$ and $B$ are zero in leading
order, making $\alpha$ indeterminate.  
Therefore,
Hara's theorem is satisfied, but at the same time $\alpha$ is not
necessarily close to zero.

\section{Summary and discussions}

Heavy-baryon chiral perturbation theory (HBChPT) \cite{hbcpt} is
considered to be an appropriate tool to study processes involving
baryons at low kinetic energies. In this papers we have studied the
constraints imposed by reparametrization invariance \cite{luke} on
heavy-baryon chiral perturbation theory (HBChPT). We have considered
the case of 3 flavors, and both the strong and the weak interaction
sectors. One feasible way to make sure the Lagrangian is
reparametrization invariant is to start from the fully relativistic
Lagrangian and take non-relativistic limit \cite{ecker}. By matching
the thus obtained Lagrangian with the most general HBChPT Lagrangian
the constraints imposed by reparametrization invariance on HBChPT can
be easily derived. In this way we established that a total of 4 terms
in the strong interaction HBChPT Lagrangian of Ref.~\cite{OP96} are
constrained, and a total number of 2 terms in the weak HBChPT
Lagrangian are constrained.  This is of importance since the total
number of free parameters available to fit the experimental data is
smaller by the constraints. For example, for the case of hyperon weak
radiative decays it is shown that as a consequence of these
constraints the leading-order parity-violating amplitudes for both
charged channels $\Sigma^+\rightarrow p+\gamma$ and $\Xi^-\rightarrow
\Sigma^-+\gamma$ are vanishing, consistent with Hara's theorem.  Since
at the same time also the parity-conserving amplitude is vanishing for
these channels, the asymmetry parameter can not be determined in
leading-order ChPT.

\section*{Acknowledgments}

This work was supported by the National Science Council of the
Republic Of China under contracts Nos.  NSC85-2112-M-007-032,
NSC85-2112-M-007-029, and NSC85-2112-M-001-026.

\end{document}